\documentclass[aps,prl,twocolumn,superscriptaddress,floatfix,letter,nolongbibliography]{revtex4-2}

\usepackage{epsfig,amsmath,amssymb,color,amsfonts,physics,microtype}
\usepackage[bookmarks=true,colorlinks,citecolor=red]{hyperref}
\usepackage{dsfont}
\usepackage{wrapfig}
\usepackage{tikz}
\usepackage{physics}
\usepackage{mathrsfs}
\usepackage[export]{adjustbox}
\usepackage{soul}
\usepackage{comment}
\usepackage{graphicx}
\usepackage{bm}
\usepackage{float}

\newcommand{\qb}{\bar{q}}
\newcommand{\q}{\bar{q}}

\begin{document}

\title{Boolean SK model}
\author{Linda Albanese}
\affiliation{Department of Mathematics and Physics, University of Salento, Italy}
\affiliation{Istituto Nazionale di Fisica Nucleare, Sezione di Lecce, Italy}
\author{Andrea Alessandrelli}
\affiliation{Department of Computer Science, University of Pisa, Italy}
\affiliation{Istituto Nazionale di Fisica Nucleare, Sezione di Lecce, Italy}

\begin{abstract}
For over half a century, statistical mechanics of spin glasses played as a paradigm  to model and interpret disparate phenomena, ranging from quantitative biology to computer science. However, despite the extensive body of research in this area, there is still a notable lack of studies addressing the replacement of Ising spins with Boolean spins: as the latter play as bits in Machine Learning, this gap to fill is now mandatory. Purpose of this paper is to address this study by focusing on the mean field assumption,  providing a comprehensive description of the results pertaining to these networks, referred to as the Boolean SK model due to their close relationship with the SK one. We provide a comprehensive framework for this model by employing Guerra interpolation: the thermodynamic limit, the replica symmetric and the broken replica free energy expressions are derived. Further, we inspect the onset of the replica symmetry breaking -i.e., the de Almeida-Thouless line- and derive  Ghirlanda-Guerra fluctuations. All theoretical findings are corroborated by numerical inspections and both highlight crucial differences in the network's behavior if compared with the Ising SK model: as the temperature is lowered, no phase transitions are evidenced and the model continuously moves from a random (ergodic) behavior to a disordered (glassy) phase. 
\end{abstract}

\maketitle

Spin glass models are subject of numerous studies in statistical mechanics since the 70s. These are network of Ising spins, variables which can be equal to $-1$ or $+1$, connected in pairs. First pivotal modelling is due to Edward and Anderson \cite{edwards1975theory}, who investigated spins arranged in a $d-$dimensional lattice with nearest neighbor interactions. The global behavior of the system is accounted by a single order parameter, the two replica overlap, which is the normalized dot product between two copies of the system (usually denoted as \textit{replicas}) sharing the same couplings. This study is prohibitively difficult to be addressed and, soon later, Sherrington and Kirkpatrick (SK) \cite{sherrington1975solvable} focused on the simpler mean-field approximation (where all the spins are linked to each other). SK immediately pointed out that their resolution of the model, based on the assumption -- known as \textit{Replica Symmetric} (RS) -- that the order parameter's probability distribution is a Dirac's delta peaked in the equilibrium value, does not give the correct solution of the model as the related entropy becomes negative in the zero temperature limit. Parisi \cite{MPV} figured out what the suitable distribution  was correct for the overlap, which giving rise to the phenomenon of \textit{Replica Symmetry Breaking} (RSB). \\ Since then, spin glass Literature has been widely developed  and it is worth mentioning some of the milestones for the purposes of this work. Keeping the focus on the SK model, De Almeida and Thouless (AT) \cite{de1978stability} wonder if a critical line between the RS and RSB solutions exists and found it via replicas method. Guerra, instead, in \cite{guerra_broken} developed a new mathematical method to recover the expression of the quenched statistical pressure of the SK model, both in RS and under finite steps of RSB. Further, together with Ghirlanda \cite{ghirlanda1998general}, they have shown how the absence of self-averaging for the overlap gives rise to fluctuations compatible with Parisi solution and together with Toninelli \cite{guerra2002thermodynamic} they have shown how to prove the existence of the asymptotic limit  of all the thermodynamic observables, ensuring the whole theory to be well defined.
\newline
However, all these results have been made for Ising spins, namely for variables which can assume $-1, +1$ values. One can wonder what happens if we replace them with Boolean spin, i.e. with $0,1$ values. The question arises naturally nowadays as in Machine Learning and Artificial Intelligence \cite{erba2024statistical, hinton2002training, lecun2015deep}  we can assimilate a spin with a computational bit. Therefore, in order to understand difference and similarities, the purpose of this work is to extend the aforementioned results from the Ising SK model to the corresponding  Boolean  counterpart, which we name by no surprise \textit{Boolean SK model}. 

\par\medskip
Once introduced the model, we prove the existence of its thermodynamic limit and we derive, through Guerra's interpolation, the expression for the quenched statistical pressure under the RS and first step of RSB (1RSB) assumptions. We  then obtain the Ghirlanda-Guerra identities and the AT line to detect the onset of replica symmetry breaking. Differences among the Ising and the Boolean SK models are highlighted and    analytical results are consistently corroborated by numerical findings.

\par\medskip
The model is ruled by an Hamiltonian defined as
\begin{align}
    \mathcal{H}_N(\bm s \vert \bm J, H) &= -\dfrac{1}{2\sqrt{N}}\sum_{i,j=1}^N J_{ij} s_i s_j  - H \sum_{i=1}^N s_i,
    \label{eq:Hamiltonian}
\end{align}

where $s_i \in \{0, 1\}$ for $i=1, \hdots ,N$, $J_{ij} \sim \mathcal{N}(0,1)$, $i,j=1, \hdots, N$ are i.i.d. standard Gaussian variables and $H\in \mathbb{R}$ is the strength of the external field.

To achieve quantitatively the control of the network behavior, we work out a statistical-mechanics investigation and we start by introducing the Boltzmann-Gibbs measure 
\small
\begin{equation}
{B}_N(\bm{s}|\bm{J},H)=\frac{e^{-\beta \mathcal{H}_N(\bm s \vert \bm J,H )}}{Z_N(\bm J,H)},\  Z_N(\bm J,H)=\sum_{\{\bm s\}}e^{-\beta \mathcal{H}_N(\bm s \vert \bm J,H )},
\end{equation}
\normalsize
where $Z(\bm J,H)$ is the partition function associated to the Hamiltonian \eqref{eq:Hamiltonian}. The parameter $\beta:=\dfrac{1}{T} \in \mathbb{R}^+$, also named as inverse temperature, tunes the distribution broadness, that is why we will call it \textit{control parameter}. 

Further, to describe the collective behavior of the system, we introduce the following order parameter  and we  refer to it as \textit{complete two replica overlap} 
\begin{equation}
    Q_{ab}= 4\left(q_{ab}-q_{aa}\right)+1
    \label{eq:total_overlap}
\end{equation}
where $a$ and $b$ denote different replicas and  
\begin{align}
    q_{ab}= \dfrac{1}{N} \sum_{i=1}^N s^{(a)}_i s^{(b)}_i,
    \label{eq:bool_overlap}
\end{align}
which is a straightforward adaptation in the case of Boolean spins of the well known  \textit{two replica overlap}  \cite{edwards1975theory, MPV}. Indeed the latter alone (i.e. Eq. \eqref{eq:bool_overlap}) is inadequate to tackle the present model, where spin-flip symmetry is broken by definition.  This is  why we introduce the generalization presented in Eq. \eqref{eq:total_overlap}. Specifically, when $Q_{ab}=1$, the two replicas $a$ and $b$ are perfectly aligned (and, oversimplifying, we can interpret their related spin configuration as representing the maximum possible order). Conversely, when $Q_{ab}= 0$, the two replicas are almost orthogonal (and, again, we can view this as representing the maximum possible disorder in the system, often referred to as the \textit{ergodic region}).


Pivotal for a statistical mechanical analysis is the study of the \textit{thermodynamic limit ($N\to \infty$) quenched statistical pressure}, defined as
\begin{align}
\label{eq:Af}
    {A}(\beta, H)&= \lim_{N \to +\infty}{A}_N(\beta,H) \notag \\
    &= \lim_{N \to +\infty} \dfrac{1}{N} \mathbb{E}_{\bm J} \log Z_N(\bm J,H)\notag \\
    &=-\lim_{N \to +\infty}\beta f_N(\beta,H)=-\beta f(\beta,H),
\end{align}
where $\mathbb{E}_{\bm J}$ is the expectation with respect to all the quenched coupling  $J_{ij}$, whereas $f_N$ is known as the quenched free energy at size $N$. 

\par\medskip
Inspired by Guerra and Toninelli's work in \cite{guerra2002thermodynamic} and leveraging Fekete's Lemma \cite{fekete1923verteilung}, we have proved \footnote{A detailed proof can be found in the Supplementary Material [URL will be inserted by the publisher].} that the thermodynamic limit of the quenched statistical pressure of the Boolean SK model \eqref{eq:Af} exists and it holds that
\begin{align}
    A(\beta,H)=\lim_{N \to +\infty} A_{N}(\beta,H) = \sup_N A_{N}(\beta,H).
\end{align}

\begin{figure}[t]
    \centering
    \includegraphics[width=8.65cm]{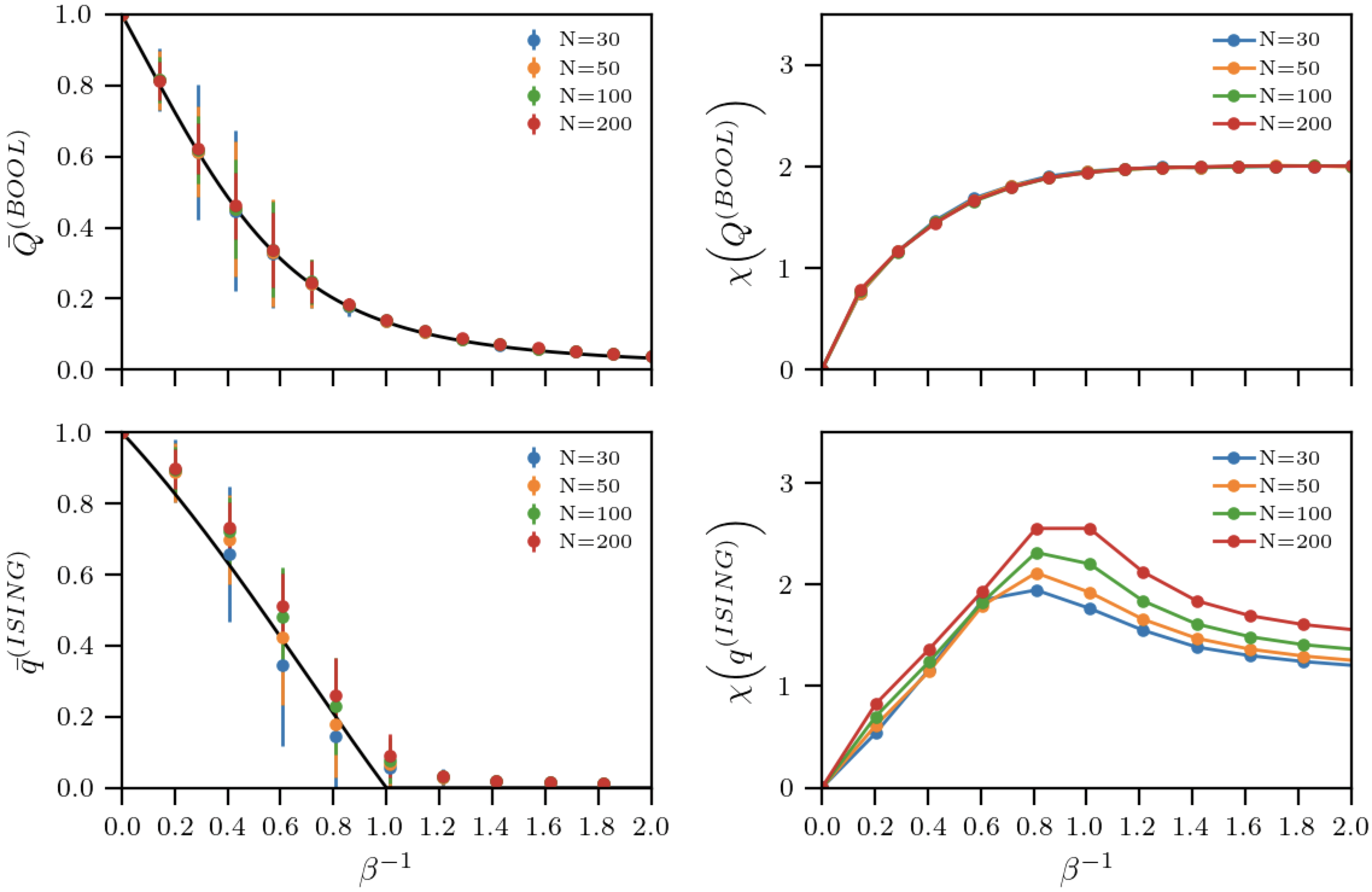}
    \caption{(Upper Panels): We present both analytical and numerical results for the Boolean SK model. In the left panel, we show the equilibrium value of the two replica complete overlap. The solid black line is obtained by numerically solving the self-consistency equations \eqref{eq:SCE} for $\bar{q}$ and $\bar{M}$, and substituting these values into \eqref{eq:total_overlap}. The colored dots represent the results of Monte Carlo (MC) simulations for different network sizes ($N$), as indicated in the legend. We emphasize that, as shown in the upper right panel, the susceptibility does not diverge as in the standard SK model with Ising-like spins, and there is no phase transition in the finite temperature case.
\\
(Lower Panels): We show the corresponding results of the upper panels for the Ising SK model at $H=0$. In the left panel, we report the MC simulation for the two replica overlap and the corresponding numerical solution of the well-known self-consistency equation, namely $\bar{q}^{(ISING)} = \mathbb{E}_x[\tanh^2(x\beta\sqrt{\bar{q}^{(ISING)}})]$. In the right panel, we plot the corresponding susceptibility. Here, one can observe a second-order phase transition at the critical point $\beta^{-1} = 1$ in the thermodynamic limit, where the susceptibility exhibits a peak as the network sizes increases.
}
    \label{fig:confronto}
\end{figure}

Since the quenched statistical pressure exists, we search for its explicit expression in terms of the complete two-replica overlap: under the RS \textit{ansatz}, the probability distribution  $\mathbb{P}(X)$ for the order parameter $X$ -- whose expected value we call $\bar{X}$ -- displays a Dirac's delta centered at $\bar{X}$ and reads as 
\begin{align}
    \mathbb{P}(X)=&\lim_{N \to + \infty} \mathbb{E}_J \mathbb{P}_{N, J}(X)= \delta(X-\bar X).
\end{align}
Therefore, in our case
\begin{align}
    \mathbb{P}(q_{ab})=&\lim_{N \to + \infty} \mathbb{E}_J \mathbb{P}_{N, J}(q_{ab})\notag \\
    &= \delta_{ab}\delta(q_{ab}-\bar M) + (1-\delta_{ab})\delta(q_{ab}-\bar q),
\end{align}
where the expected values are $\bar q$, when the two replicas are different, and $\bar M$, where the two replicas coincide. Note that, as there is no spin-flip symmetry in this model,  $\bar M$ accounts for its intrinsic magnetization.
Using \textit{Guerra's interpolation} \cite{guerra_broken} we obtained a first approximation of the quenched statistical pressure for the Boolean SK model in the thermodynamic limit, that is
\begin{equation}
\begin{array}{lll}
    A^{(RS)}(\beta,H) &&=\beta\dfrac{H}{2}+\dfrac{\beta^2}{8}\left(\bar{M}-\bar{q}\right)-\dfrac{\beta^2}{8}\left[\bar{M}^2 -\bar{q}^2\right] \\\\
    &&+\log 2+\mathbb{E}_{x}\log\left[\cosh g(\bar{M}, \qb \vert \beta, H,x)\right],
\end{array}
\label{eq:press_stat_finale}
\end{equation}
where $x$ is a standard Gaussian variable, $$g(\bar{M}, \qb \vert \beta, H,x)=\dfrac{\beta^2}{8}\left(\bar{M}-\bar{q}\right)+x\sqrt{\dfrac{\beta^2}{8}\bar{q}} +\beta\dfrac{H}{2}$$ and the order parameters $\bar M$ and $\bar q$ fulfill the following self-consistency equations
\begin{equation}
\begin{array}{lll}
    \bar{M} &=& \dfrac{1}{2}\left[1+\mathbb{E}_{x}\tanh g(\bar{M}, \qb \vert \beta, H,x) \right],
    \\\\
    \bar{q} &=& \bar{M} -\dfrac{1}{4}\left[1-\mathbb{E}_{x}\tanh^2 g(\bar{M}, \qb \vert \beta, H,x) \right].
\end{array}
\label{eq:SCE}
\end{equation}
The examination of Eqs. \eqref{eq:total_overlap} and  \eqref{eq:SCE} offers a quantitative understanding of the system's behavior as $\beta$ changes.

Numerically solving \eqref{eq:SCE} and then replacing the result in \eqref{eq:total_overlap} allows us to obtain a full picture of the behavior of the Boolean network under analysis and to compare it with its Ising counterpart, which is governed by the following self-consistency equation:
\begin{equation}
    \label{eq:SCE_Ising}
    \bar{q}^{(ISING)}=\mathbb{E}_x\tanh^2(x\beta\sqrt{\bar{q}^{(ISING)}}).
\end{equation}
As illustrated in Fig. \ref{fig:confronto}, both the Ising and Boolean models exhibit maximal ordering as the temperature approaches $0$ and maximal disorder as the temperature tends to $+\infty$. However, unlike its Ising counterpart, which undergoes a second-order phase transition at the critical temperature $\beta^{-1}=1$, the Boolean SK model does not exhibit any phase transition at finite temperature, neither first nor second-order. This is evident from the panels in Fig. \ref{fig:confronto}, where the Ising model shows a different behaviour and, in its susceptibility, there is growing peak at $\beta^{-1}=1$ as the system size increases in Monte Carlo (MC) simulations, while no such peak is observed for the Boolean model.



\par\medskip
The analysis carried out for the Boolean SK model has been addressed within a RS scheme. However, since we have dealt with a particular realization of spin-glasses, such symmetry is expected to be broken in the low temperature limit, where the system develops a multitude of degenerate states. This behaviour, along with the imprecision of the RS \textit{ansatz} (\textit{RS instability}), is easy to check in the low-temperature region by looking at the RS-predicted entropy of the system.
Starting from the RS expression for the quenched statistical pressure, the corresponding entropy can be derived using the general relation $s(\beta,H) =  \beta^{2}\partial_{\beta} f(\beta,H)$. This leads to the following result, supposing $H=0$ without loss of generality,
\begin{align}
    s^{\textnormal{(RS)}}(\beta, H=0)&=\log 2 +\dfrac{\beta^2}{8} (\bar{M}-\q)-\dfrac{3\beta^2}{8} (\bar{M}^2 - \q^2) \notag \\
    &-\mathbb{E}_{x}\log\left[\cosh g(\bar{M}, \qb \vert \beta, x, H=0)\right],
    \label{eq:entropiaRS}
\end{align}
where $\bar M$ and $\bar{q}$ fulfill the self-consistency equations \eqref{eq:SCE}.

\begin{figure}[t]
    \centering
    \includegraphics[width=8 cm]{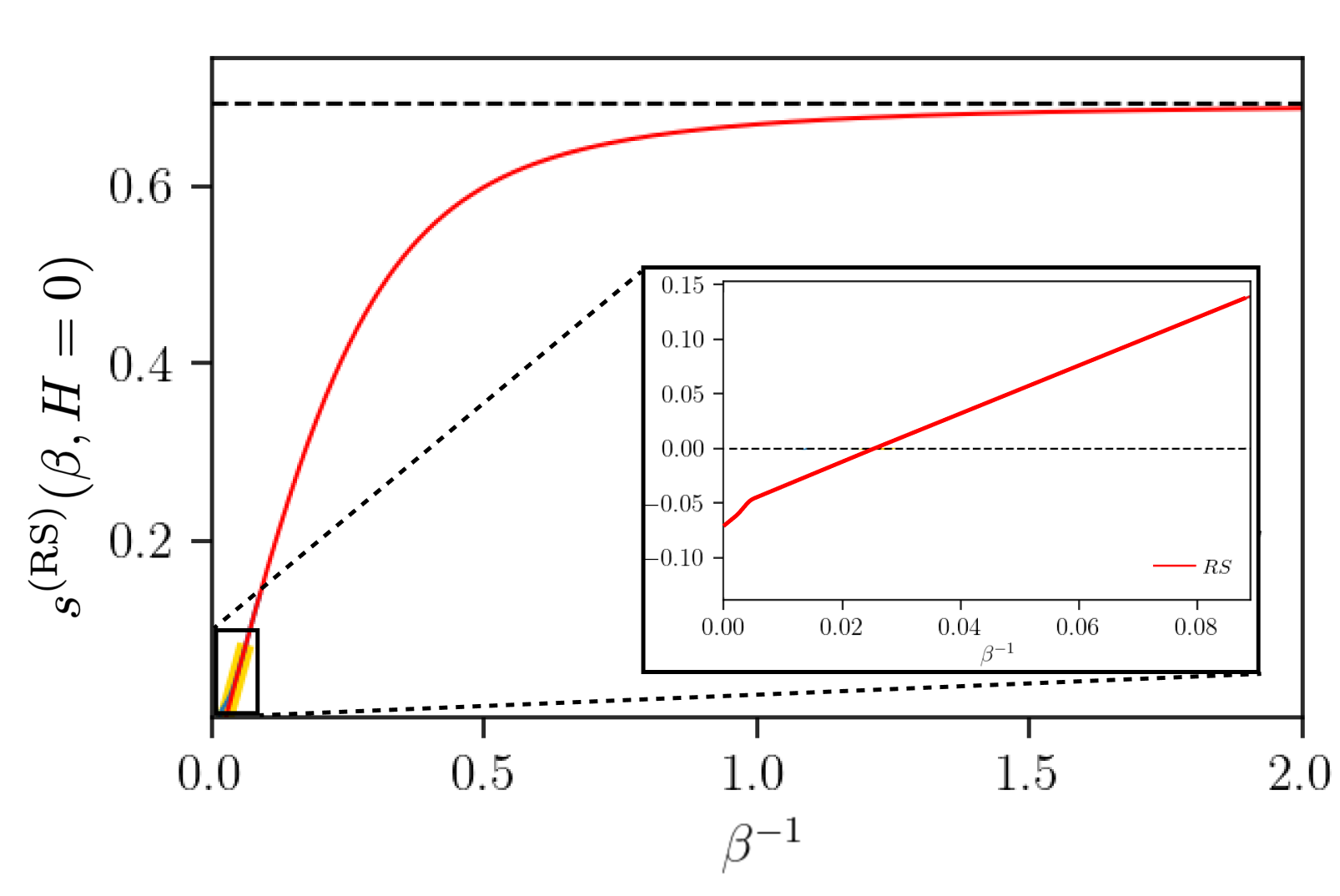}
    \caption{Using the results obtained from the numerical solution of the self-consistency equations \eqref{eq:SCE} under the RS assumption, we compute the RS entropy of the model with $H=0$ \eqref{eq:entropiaRS}. As expected, for large values of the temperature $\beta^{-1} \to \infty$, the RS entropy saturates to a value of $\log 2$, as it should. However, as shown in the zoomed-in plot, for small temperatures ($\beta^{-1} \leq 0.02$), under RS assumption, we observe negative entropy values, which is a physically meaningless result for a discrete system like the one under analysis.}
    \label{fig:entr_RS_neg}
\end{figure}

As one can see from Fig. \ref{fig:entr_RS_neg}, the RS entropy appears to be negative near zero temperature. This result is physically forbidden, as the entropy of a discrete system must always be positive or zero. This indicates that the RS assumption we employed for the distribution of the order parameter may not be the most appropriate. Further evidence of this can be seen by scanning the GG \cite{ghirlanda1998general} identity we derived for this model, where we find the following relation holds:
\begin{align}
     \left\langle \left(q_{12}^2-q_{13}^2\right)^2\right\rangle+\Big(\left\langle q_{11}^2\right\rangle-\left\langle q_{12}^2\right\rangle\Big)^2=\left\langle \left(q_{11}^2-q_{13}^2\right)^2\right\rangle,
     \label{eq:GGNew}
\end{align}
which will be trivially true if all the replica overlaps are delta-peaked at $\bar{q}$, when the two replicas are different, and at $\bar M$, when the two replicas coincide.

Following the approach already explored for the standard SK model \cite{MPV, guerra_broken}, we now introduce a more sophisticated ansatz: the RSB framework. This approach is justified by the fact that, as demonstrated in \cite{MPV, ghirlanda1998general}
, the behavior and distribution of the order parameters under the RSB assumption in glassy models also lead to relations similar to the one presented in \eqref{eq:GGNew}.

\par\medskip
To carry out the analysis at this level of symmetry breaking, we adopt the standard 1RSB assumption \cite{MPV}, which is stated as follows: the probability distribution of the order parameter $q_{ab}$, when the two replicas differ, exhibits two peaks $\q_1$ and $\q_2$ whose concentration is ruled by a parameter $\theta \in [0,1]$:
\begin{align}
    \mathbb{P}(q_{ab})=& \lim_{N \to +\infty} \mathbb{E}_{\bm J}\mathbb{P}_{N, J}(q_{ab})= \delta_{ab} \delta(q_{ab}- \bar M) \notag \\
    &+ (1-\delta_{ab})\left((1-\theta)\delta(q_{ab} -\q_2) + \theta \delta(q_{ab}- \q_1) \right).
\end{align}

Following the path from Guerra in \cite{guerra_broken} using the 1RSB assumption we get that, in the thermodynamic limit, the quenched statistical pressure of the Boolean SK model reads as
we get that, in the thermodynamic limit, the quenched statistical pressure of the Boolean SK model reads as
\begin{equation}
\begin{array}{lll}
     &&\mathcal{A}_{\theta}^{(1RSB)}(\beta, H) = \dfrac{1}{\theta}\mathbb{E}_1\log\mathbb{E}_2 W(\beta, H, \bm Y, \theta)\notag \\\\
     &&+\log 2+\dfrac{\beta H}{2} -\dfrac{\beta^2}{8}\Big(\bar{M}^2-\bar{q}^2_2\Big) -\theta\dfrac{\beta^2}{8}\Big(\bar{q}^2_2-\bar{q}_1^2\Big) 
\end{array}
\label{eq:A1-RSB}
\end{equation}
where
\small
\begin{equation}
\begin{array}{lll}
     g(\beta,H,\bm Y) =&& \dfrac{\beta H}{2} +\dfrac{\beta^2}{8}(\bar{M}-\bar{q}_2) + Y^{(1)}\sqrt{\dfrac{\beta^2}{8}\bar{q}_1}\\&&+ Y^{(2)}\sqrt{\dfrac{\beta^2}{8}(\bar{q}_2-\bar{q}_1)},
     \\\\
    W(\beta,H,\bm Y, \theta)=&&  e^{\theta g(\beta, H, \bm Y)}\cosh^\theta\Big[g(\beta, H,\bm Y)\Big]
\end{array}
\end{equation}
\normalsize
and the order parameters must fulfill the following self-consistency equations
\small
\begin{equation}
\begin{array}{lll}
     \bar{M}= \dfrac{1}{2}\left[1+\mathbb{E}_1\left(\dfrac{\mathbb{E}_2 W(\beta, H, \bm Y, \theta) \tanh g(\beta, H, \bm Y) }{\mathbb{E}_2 W(\beta,H, \bm Y, \theta)}\right)\right],
     \\\\
     \bar{q}_2= \Delta+\dfrac{1}{4}\mathbb{E}_1\left\{\dfrac{\mathbb{E}_2 W(\beta,H,\bm Y, \theta) \tanh^2 g(\beta,H,\bm Y)}{\mathbb{E}_2 W(\beta,H,\bm Y, \theta)}\right\},
     \\\\
     \bar{q}_1=\Delta+\dfrac{1}{4}\mathbb{E}_1\left\{\dfrac{\mathbb{E}_2 W(\beta, H, \bm Y, \theta) \tanh g(\beta,H,\bm Y) }{\mathbb{E}_2 W(\beta,H,\bm Y, \theta) }\right\}^2.
\end{array}
\label{eq:SCE_1-RSB}
\end{equation}
\normalsize
with $\Delta = \bar{M} -\frac{1}{4}$ and where
\small
\begin{equation}
\begin{array}{lll}
     &&\theta^2=\dfrac{8}{\beta^2(\bar{q}_1^2-\bar{q}_2^2)}\mathbb{E}_1\Bigg\{\log\mathbb{E}_2 W(\beta,H,\bm Y, \theta)
     \\\\
&&\left.\hspace{-0.5cm}-\theta\dfrac{\mathbb{E}_2 \left[W(\beta,H,\bm Y, \theta)\Big(g(\beta,H,\bm Y) +\log\cosh[g(\beta,H,\bm Y)]\Big)\right]}{\mathbb{E}_2  W(\beta,H,\bm Y, \theta)}\right\}. 
\end{array}   
\end{equation}
\normalsize

\begin{figure}[t]
    \centering
    \includegraphics[width=8 cm]{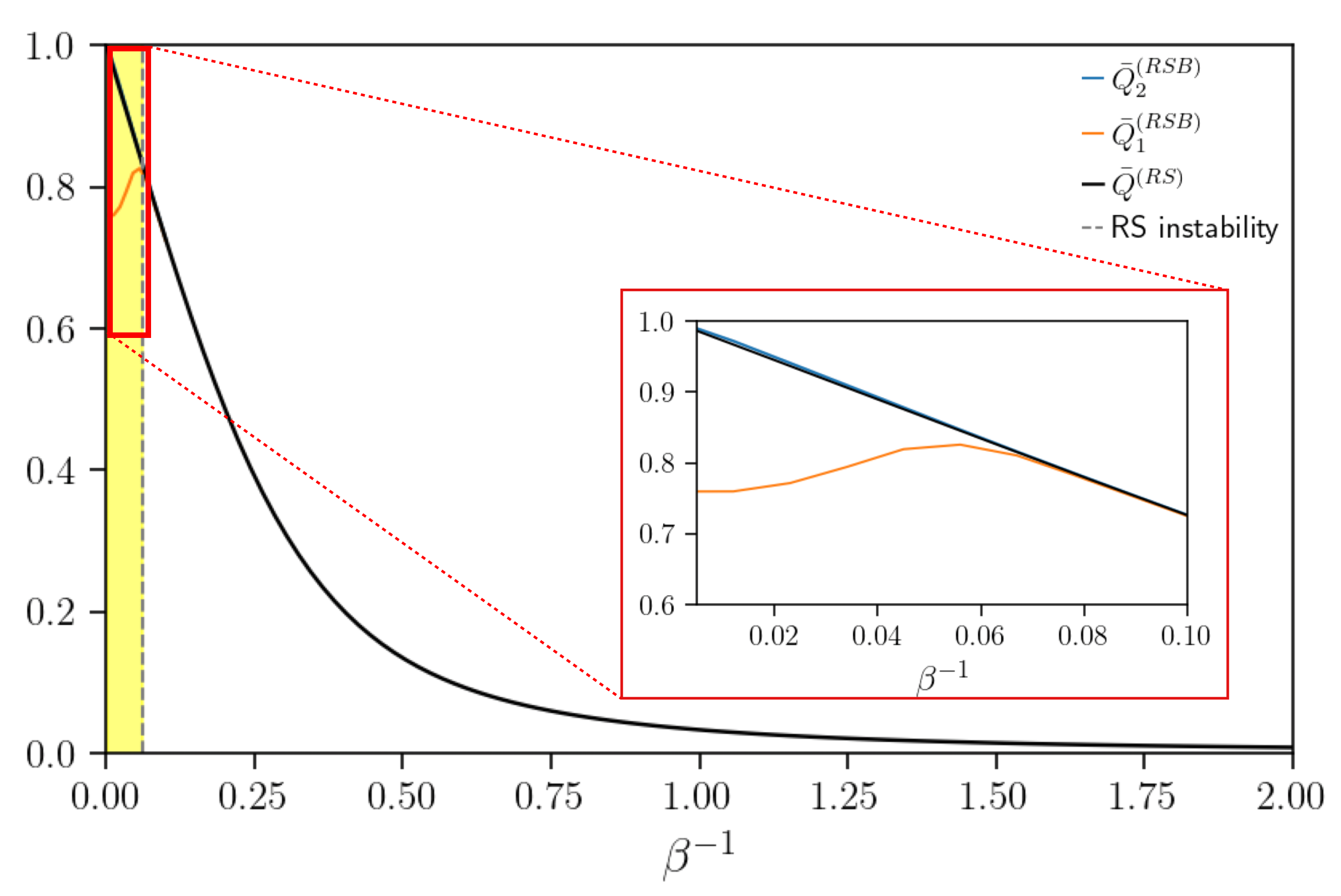}
    \caption{Representation of the two 1RSB peaks of the complete two replica overlap $\bar{Q}_a^{(RSB)} = 4 \left( \q_a -\bar{M} \right) + 1$, $a=1, 2$, computed numerically via  \eqref{eq:SCE_1-RSB} and the RS mean value $\bar{Q}^{(RS)}$ using \eqref{eq:SCE}. The region highlighted in yellow is zoomed in on the red square, representing the instability of the RS assumption as indicated by the AT line \eqref{eq:AT_RS}. We observe that the RS and 1RSB solutions differ only when $\bar{Q}_1^{(RSB)}$ and $\bar{Q}^{(RSB)}_2$ begin to diverge, which occurs at low values of $\beta^{-1}$, where the RS solution starts to become unstable, as expected.
}
    \label{fig:self1-RSB_plot}
\end{figure}

\par\medskip
We stress that if we put $\theta=0$ or $\theta=1$ in the expression of the 1RSB quenched statistical pressure \eqref{eq:A1-RSB} we come back to the one previously found in RS assumption \eqref{eq:press_stat_finale}, namely
$$
\mathcal{A}^{(RS)}(\beta,H)=\mathcal{A}_{\theta=0}^{(1RSB)}(\beta,H)=\mathcal{A}_{\theta=1}^{(1RSB)}(\beta,H).
$$
Therefore, another important question we want to address here is for which value of the control parameter (i.e. $\beta$) the RS instability emerges. Our purpose is then to prove that for values of $\theta$ close but away from $0$ or $1$, the 1RSB expression of the quenched statistical pressure is greater than the RS one, i.e. 
\begin{equation*}
    \begin{array}{lll}
         \mathcal{A}^{(RS)}(\beta,H)<\mathcal{A}_{\theta=0}^{(1RSB)}(\beta,H),
         \\\\
         \mathcal{A}^{(RS)}(\beta,H)<\mathcal{A}_{\theta=1}^{(1RSB)}(\beta,H),
    \end{array}
\end{equation*}
below a critical line in the parameter space $\beta, \ H$. In a recent work \cite{albanese2023almeida}, we have devised a simple and rigorous method to detect the onset of the instability of RS theories in neural networks. This has been done via the computation of the AT line, introduced for the first time by the two physicists in \cite{de1978stability} for SK model. 

We put $H=0$ for mathematical convenience. The RS assumption becomes unstable when $\beta$ fulfills the following expression 
\begin{align}
     1-\dfrac{\beta^2}{32} \mathbb{E}_x \left[{\textnormal{sech}^4 \left(\dfrac{\beta^2}{8}(\bar{M} - \q) + x\sqrt{\dfrac{\beta^2}{8} \q} \right)}\right] < 0,
     \label{eq:AT_RS}
\end{align}
where $\bar{M}$ and $\bar{q}$ fulfill the RS self-consistency equations \eqref{eq:SCE}. One can numerically verify (see Fig. \ref{fig:self1-RSB_plot}) that the RS \eqref{eq:SCE} and 1RSB \eqref{eq:SCE_1-RSB} solutions begin to exhibit different behavior precisely in the region predicted by \eqref{eq:AT_RS}. 

\begin{figure}[t]
    \centering
    \includegraphics[width=8 cm]{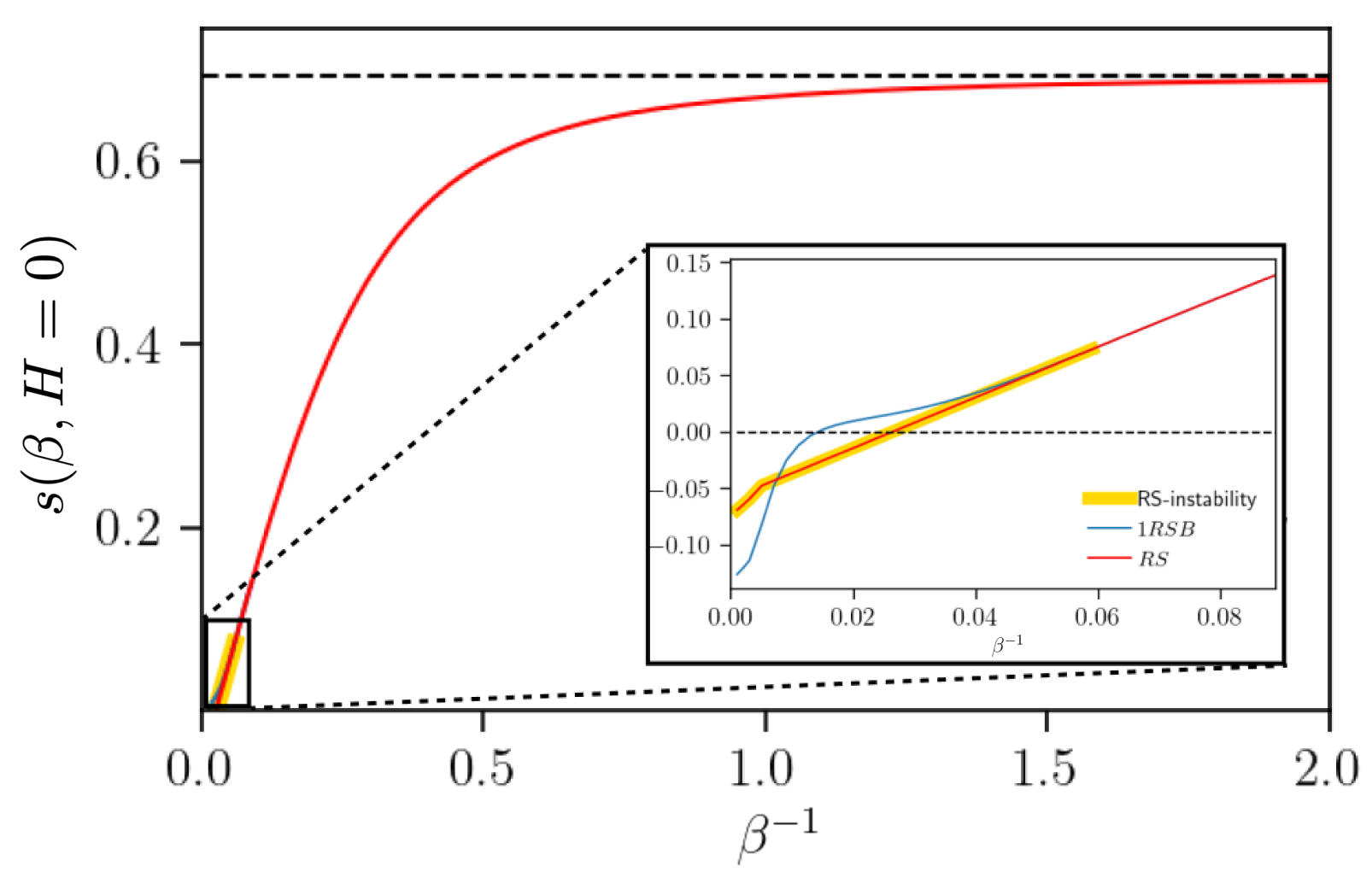}
    \caption{Comparison between the entropy computed under the RS (red) and 1RSB (blue) assumptions with respect to the temperature $\beta^{-1}$. The yellow line, computed via the AT line \eqref{eq:AT_RS}, marks the temperature region where the RS solution becomes unstable —- specifically, the region where the RS and 1RSB solutions start to diverge. A zoomed-in view is provided to highlight the differences between the two assumptions at low temperatures. Indeed, in the small temperature region, the 1RSB assumption extends the validity of the solution, as the entropy becomes negative at lower temperatures compared to the RS solution. However, even under the 1RSB assumption, the entropy still becomes negative as $\beta^{-1}\to 0$, though at smaller temperature values compared to the RS solution. We emphasize that, at values of temperature approaching zero, numerical calculations of 1RSB entropy experience significant instability. Nonetheless, this region would be physically inadmissible regardless, as entropy assumes negative values there.}
    \label{fig:entropy_plot}
\end{figure}

As for the 1RSB entropy, the expression can be computed using the same relation as in the RS approximation. As shown in Fig.  \ref{fig:entropy_plot}, where we also juxtapose the results from the AT line, which accurately predicts the transition from RS to the 1RSB approximation, the solution of the latter reduces the interval in which the entropy is negative, thereby improving the range of temperatures where our new approximation is valid. 
However, as we can see, the 1RSB assumption also presents a region near $\beta\to\infty$ where it predicts a negative entropy, which corresponds to a non-physical region. This highlights once again that this solution is not the correct one. The correct solution is likely to be a KRSB steps approximation or even, as in the standard SK model \cite{MPV}, a full RSB ($\infty$RSB) solution.


\section{Discussions}

The Boolean SK model, while defined similarly to the Ising SK model, differs in several key aspects. Notably, it features an internal magnetization that is independent of any external field (due to the break of the spin-flip symmetry), compensating for the absence of randomness in the interactions. As a consequence, the \textit{complete two replica overlap}, which retains all the descriptive properties of the standard two replica overlap used in the Ising SK model and provides information about the similarity between two replicas, does not exhibit the characteristic phase transition seen in the Ising counterpart.

Once proved the existence of the asymptotic values for our theory, we first adopt the RS assumption and  prove it to be wrong in the low-temperature limit (an explicit computation of the RS entropy shows that it becomes negative in this limit). To address this, we investigate the 1RSB à la Parisi, alongside the study of the AT line.

The AT line accurately predicts the region of the RS instability, where the RS ansatz becomes physically meaningless. In this context, the structure of 1RSB integrates naturally, and the results obtained provide an improved approximation compared to the simpler RS ansatz.

Additionally, it is particularly intriguing how the GG identities are modified in this model, acquiring a distinct structure from that of the Ising SK network due to the presence of the internal magnetization.

All numerical findings are consistent with the previously obtained analytical results.

This work serves as a precursor to a broader series of studies. It will be fascinating to explore how these changes affect other types of networks, such as neural networks \cite{Hopfield}, when considering Boolean spins. Such exploration could hold great significance in the context of eXplainable AI \cite{xu2019explainable}.


\par\medskip
\acknowledgments
The authors acknowledge Adriano Barra for very useful discussions. \\
L.A. acknowledges the PRIN grant {\em Statistical Mechanics of Learning Machines} n. 20229T9EAT for financial support.\\
A.A. acknowledges INdAM (Istituto Nazionale d'Alta Matematica) and UniSalento for support via PhD-AI. \\
The authors are members of the group GNFM of INdAM which is acknowledged too. 


%

\end{document}